\documentclass[12pt]{article}
\usepackage{amssymb,amsmath,graphicx}

\newcommand{\be}{\begin{equation}}
\newcommand{\ee}{\end{equation}}
\newcommand{\bea}{\begin{eqnarray}}
\newcommand{\eea}{\end{eqnarray}}
\newcommand{\beas}{\begin{eqnarray*}}
\newcommand{\eeas}{\end{eqnarray*}}

\newcommand{\phic}{\varphi}
\newcommand{\gl}{\gamma_{\lambda}}
\newcommand{\gp}{\gamma_{\varphi}}

\begin{document}
\begin{titlepage}
\begin{center}

{\Large \bf Bouncing and cyclic string gas cosmologies}

\vspace{6mm}

\renewcommand\thefootnote{\mbox{$\fnsymbol{footnote}$}}
Brian Greene${}^1$\footnote{greene@phys.columbia.edu},
Daniel Kabat${}^{1,2}$\footnote{daniel.kabat@lehman.cuny.edu} and
Stefanos Marnerides${}^1$\footnote{stefanos@phys.columbia.edu}

\vspace{4mm}

${}^1${\small \sl Institute for Strings, Cosmology and Astroparticle Physics} \\
{\small \sl and Department of Physics} \\
{\small \sl Columbia University, New York, NY 10027 USA}

\vspace{4mm}

${}^2${\small \sl Department of Physics and Astronomy} \\
{\small \sl Lehman College, City University of New York} \\
{\small \sl Bronx, NY 10468 USA}

\end{center}

\vspace{1cm}

\noindent
We show that, in the presence of a string gas, simple
higher-derivative modifications to the effective action for gravity
can lead to bouncing and cyclic cosmological models.  The
modifications bound the expansion rate and avoid singularities at
finite times.  In these models the scale factors can have long
loitering phases that solve the horizon problem.  Adding a potential
for the dilaton gives a simple realization of the pre-big bang
scenario.  Entropy production in the cyclic phase drives an eventual
transition to a radiation-dominated universe.  As a test of the
Brandenberger-Vafa scenario, we comment on the probability of
decompactifying three spatial dimensions in this class of models.

\end{titlepage}
\setcounter{footnote}{0}
\renewcommand\thefootnote{\mbox{\arabic{footnote}}}

\section{Introduction}

A fundamental question in cosmology is whether the universe has always
existed, or whether it came into being a finite time in our past.  It
could be that the age of the universe is finite; at the classical
level the singularity theorems of general relativity make such an
assumption seem unavoidable \cite{Hawking:1969sw}.  The other
possibility is that the universe has infinite age.  A number of
past-eternal models have been developed, exploiting the fact that
quantum effects or other modifications to general relativity can get
around the singularity theorems \cite{Novello:2008ra}.

String theory should ultimately provide a framework for deciding
between a moment of creation and an eternal universe.\footnote{Eternal
inflation, despite its name, is not past-eternal and cannot by
itself address this issue \cite{Borde:1993xh,Borde:2001nh}.}  On the
one hand various toy models for cosmological singularities in string
theory have been developed, and considerable effort has been devoted
to studying them, but a complete understanding is still lacking
\cite{Berkooz:2007nm}.  On the other hand several string-inspired
models for eternal cosmologies have been proposed, most notably the
ekpyrotic \cite{Khoury:2001wf} and pre-big bang \cite{Gasperini:2002bn}
scenarios, but it is not clear to what extent these proposals capture
the generic (or even allowed) behavior of string theory.

In view of this situation it is worthwhile developing additional
scenarios for eternal cosmologies in string theory.  In this paper we
consider a simple class of higher-derivative modifications to the
effective action for gravity.  These modifications have the effect of
bounding the expansion rate and limiting dilaton gradients, thereby
avoiding singularities at any finite time.  In the absence of matter the
universe would approach a de Sitter phase at early times.  But when
coupled to a gas of string winding and momentum modes the scale
factors can oscillate or bounce as functions of time.  By introducing
a dilaton potential the dilaton can be made to oscillate or bounce as
well.  Our work has several motivations.

\noindent
{\em Bouncing and cyclic cosmologies} \\
Eternal cosmologies in which the scale factors bounce or oscillate as
functions of time have been extensively studied, and within field
theory a variety of mechanisms for realizing this type of behavior
have been developed \cite{Novello:2008ra}.  Our work provides a simple
string-inspired mechanism for obtaining bouncing and cyclic cosmologies.
For other studies in this direction see \cite{Fabris:2002pm,Biswas:2005qr,Biswas:2006bs}.

\goodbreak
\noindent
{\em Pre-big bang scenario} \\
In the pre-big bang scenario the universe is assumed to begin from a
cold flat weakly-coupled initial state.  The dilaton rolls towards
strong coupling and bounces, and the universe emerges in an expanding
FRW phase \cite{Gasperini:2002bn}.  However at the level of the
two-derivative effective action the two branches of pre-big bang
cosmology cannot be smoothly connected \cite{Brustein:1994kw}.  There
are various ways around this, reviewed in section 8 of
\cite{Gasperini:2002bn}.  Our work leads to a particularly simple
realization of the pre-big bang scenario, in a manner similar to
the proposal \cite{Brandenberger:1998zs}.

\noindent
{\em Horizon problem} \\
One of the main puzzles of conventional FRW cosmology is the horizon
problem: how did causally-disconnected regions of the universe come to
be in thermal equilibrium?  Inflation explains this by postulating a
rapid growth of the scale factor at early times.  But an alternative
way to address the horizon problem is to postulate a loitering phase
in which the scale factor is roughly constant (that is, the Hubble
length diverges).  If the loitering phase lasts long enough the
universe has time to come to thermal equilibrium.  The models we
discuss can have long loitering phases, a phenomenon observed in a
similar context in \cite{Brandenberger:2001kj}.

\noindent
{\em Brandenberger-Vafa mechanism} \\
As a zeroth-order goal, one might hope that string cosmology could
account for the three large spatial dimensions we observe.  An
intriguing dynamical mechanism for obtaining three large dimensions
was proposed by Brandenberger and Vafa \cite{brandvafa}; see
\cite{Battefeld:2005av} for a review.  They imagine
the universe began at the Hagedorn temperature, with all spatial
dimensions compactified on a string-scale torus.  A gas of winding
modes keeps the universe from expanding, while a gas of momentum modes
keeps it from shrinking, until a thermal fluctuation happens to make
some number of dimensions expand.  If three or fewer dimensions expand
the winding strings should generically intersect, and if they happen
to annihilate there will be no obstacle to those dimensions
decompactifying.  But four or more dimensions should be prevented from
decompactifying, since the winding strings will generically not
intersect.  Although appealing, at the level of the two-derivative
effective action this scenario has a problem: the universe has a
singularity a finite time in the past.  Moreover the dilaton rolls
monotonically towards weak coupling, so if one waits too long strings
are unlikely to annihilate even if they do happen to intersect.  This
means there is only a small window of time for the necessary thermal
fluctuations to take place, and as a result three large dimensions are
not statistically favored \cite{winds}.  The eternal cosmologies we
discuss would seem to provide a natural setting for realizing the
Brandenberger-Vafa mechanism.  Indeed this was our original motivation
for analyzing these models.  We will find that, due to possibility of
long loitering phases, our models do not preferentially decompactify
three dimensions.

Several of the results we will obtain have antecedents in the
literature, in particular in the works
\cite{Brandenberger:1998zs,Brandenberger:2001kj}, although to our
knowledge the phenomena we will discuss have never appeared in
combination before.  An outline of this paper is as follows.  In
section \ref{sect:modified} we introduce a modified action for
Einstein-dilaton gravity which has the effect of bounding the
expansion rate and dilaton gradient.  In section \ref{sect:thermo} we
introduce matter degrees of freedom and give a preliminary discussion
of their thermodynamics.  In section \ref{sect:bouncing} we consider
the coupled gravity -- matter system and show how bouncing and cyclic
cosmologies result.  In section \ref{sect:entropy} we study string
interactions and entropy production in these cosmologies.  In section
\ref{sect:exit} we show that entropy production drives an eventual
transition to a conventional radiation-dominated cosmology.  Section
\ref{sect:BV} studies the extent to which the Brandenberger-Vafa
mechanism is operative in these models.  We conclude in section
\ref{sect:conclusions}.  In appendix \ref{sect:Econd} we discuss energy
conditions in dilaton gravity and in appendix \ref{sect:inc} we study the
fine-tuning of initial conditions required to exit the Hagedorn phase
if one uses a two-derivative effective action.

\section{A modified action\label{sect:modified}}

We consider type II string theory compactified on a torus with metric
\begin{equation}
\label{metric}
ds^2=-dt^2+\alpha^{\prime}\sum_{i=1}^{d}e^{2\lambda_i(t)}dx_i^2\hspace{8mm}
x_i \approx x_i + 2\pi\,.
\end{equation}
From now on we set $\alpha' = 1$.  Although we have in mind that all
spatial dimensions are compactified, we will allow $d$ to vary to
study the dimension dependence of our results.  At the two-derivative
level the string-frame effective action for homogeneous fields takes
the form
\begin{equation}\label{s0}
S_0=\int dt \, \Big[4\pi^2 e^{-\phic}\Big(\sum_i\dot{\lambda}_i^2-\dot{\phic}^2\Big) +
L_{\rm matter}\Big]\,.
\end{equation}
The action includes standard kinetic terms for the radii and dilaton;
we're working in terms of a shifted dilaton $\phic$, related to the
usual dilaton $\phi$ by \cite{tsetvafa}
\begin{equation}
\phic=2\phi-\sum_i\lambda_i\,.
\end{equation}
$L_{\rm matter}$ is the effective Lagrangian for matter degrees of
freedom.  In thermal equilibrium we'll identify $L_{\rm matter} = - F$
with the negative of the matter free energy.  Besides the equations of
motion which follow from this action we have the Hamiltonian
constraint that the total energy in the universe vanishes.
\begin{equation}
\label{Friedmann}
\dot{\phic}^2-\sum_i\dot{\lambda}_i^2 = {1 \over 4 \pi^2} E e^{\phic}
\end{equation}
Here $E$ is the energy in matter.  These equations are invariant
under T-duality, which acts according to
\beas
&&\lambda_i \rightarrow - \lambda_i \qquad \hbox{\rm for some $i$}\\
&&\phic,L_{\rm matter},E \qquad\quad \hbox{\rm invariant}
\eeas

Provided $E$ satisfies certain energy conditions the action $S_0$
leads to cosmologies that have initial singularities: at a finite
proper time in the past the shifted dilaton diverges and the
$\lambda_i \rightarrow \pm \infty$. These singularities have mostly
been studied in the context of pre-big bang cosmology, see for example
\cite{brus} and \cite{brus2}.  But it is tempting to speculate that
stringy effects ($\alpha'$ corrections to the effective action) will
lead to non-singular cosmologies.\footnote{For a study of $\alpha'$
corrections in string gas models see \cite{Borunda:2006fx}.}  As a model which captures this sort
of behavior, we introduce the following modified action for the metric
and dilaton.
\begin{equation}
\label{s}
S=\int dt \, \Big[8\pi^2 e^{-\phic}\Big(
\sqrt{1-\dot{\phic}^2}-\sqrt{1-\sum_i\dot{\lambda}_i^2}\,\Big) + L_{\rm matter}\Big]
\end{equation}
Again in equilibrium we'll identify $L_{\rm matter} = - F$ with the
negative of the matter free energy.

There are several motivations for writing down this action.  As a
simple way to think about it, note that in the action (\ref{s0}) both
$\phic$ and $\lambda_i$ appear as non-relativistic particles of mass
$8 \pi^2 e^{-\phic}$ (although $\phic$ has a wrong-sign kinetic
energy).  In going from (\ref{s0}) to (\ref{s}) we have promoted
$\phic$ and $\lambda_i$ to become relativistic particles of the same
mass.  This clearly bounds their velocities,
\[
\dot{\phic}^2 < 1 \qquad\quad \sum_i \dot{\lambda}_i^2 < 1
\]
which has the desired effect of ruling out singularities at any finite
proper time. In this sense the action we have written down
incorporates a ``limiting curvature hypothesis'' in a manner similar
to \cite{Brandenberger:1998zs,Mukhanov:1991zn}.  It's also amusing to
note the resemblance of $S$ to the DBI action for open strings, which
is related by T-duality to the action for a relativistic particle
\cite{Leigh:1989jq}.  Finally we note that with the conventional two-derivative
action (\ref{s0}), some fine-tuning of initial conditions is required to
exit an initial Hagedorn phase.  We discuss this in appendix \ref{sect:inc}.
With the modified action (\ref{s}), this difficulty is avoided.

For simplicity we specialize to a square torus with all $\lambda_i = \lambda$.
Then the equations of motion which follow from $S$ are
\begin{equation}\begin{aligned}
\label{eqmot}
\dot{\gp}&=\dot{\phic}(\gp-\gl^{-1})+\frac{1}{8\pi^2}\dot{\phic}e^{\phic}P_{\phic}\\
\dot{\gl}&=\dot{\phic}(\gl-\gl^{-1})+\frac{1}{8\pi^2}d\dot{\lambda}e^{\phic}P_\lambda
\end{aligned}\end{equation}
Here $P_\phic = {\partial F \over \partial \phic}$ is the force on the
dilaton and $P_\lambda = - {1 \over d} {\partial F \over \partial
\lambda}$ is the pressure (or more accurately, the pressure times
the volume of the torus).\footnote{These are derivatives at fixed
temperature.  Entropy will be conserved, until we consider
out-of-equilibrium processes in section \ref{sect:entropy}, so it's perhaps
more appropriate to write $P_\lambda = - {1 \over d} \left({\partial E
\over \partial \lambda}\right)_S$ as a
derivative at fixed entropy.}  We have defined the relativistic factors
\begin{equation}
\gp=\frac{1}{\sqrt{1-\dot{\phic}^2}}\hspace{1cm}\gl=\frac{1}{\sqrt{1-d\dot{\lambda}^2}}.
\end{equation}
The Hamiltonian constraint (Friedmann equation) is
\begin{equation}\label{encon}
\gp-\gl=\frac{1}{8\pi^2}Ee^{\phic}
\end{equation}
where $E$ is the matter energy.  Note that the positive energy region is
$|\dot{\phic}|\geq\sqrt{d}|\dot{\lambda}|$, just as in lowest order dilaton
gravity.

To get oriented, consider a simple equation of state $P_{\lambda}=wE$,
$w$ constant, with $P_{\phic}=0$. Of particular interest are the cases
$w=0$, $w=1/d$ and $w=-1/d$ which correspond to a Hagedorn era, a
radiation dominated era and a winding mode dominated era respectively.
One can get an idea of how the system evolves by writing equations for
$\ddot{\phic}(\dot{\phic},\dot{\lambda})$ and
$\ddot{\lambda}(\dot{\phic},\dot{\lambda})$ and studying the phase
space flow. Using the above equation of state, and substituting the
Hamiltonian constraint in the equation for $\lambda$ we can write
\be \label{constw}
\begin{aligned}
&\ddot{\phic}=(1-\dot{\phic}^2)(1-\gl^{-1}\gp^{-1})\\
&\ddot{\lambda}=(1-d\dot{\lambda}^2)(\dot{\phic}\dot{\lambda}-w(1-\gl^{-1}\gp))
\end{aligned}
\ee

\begin{figure}[htp]
\centering
\includegraphics[scale=0.8,viewport=20 100 450 720,clip]{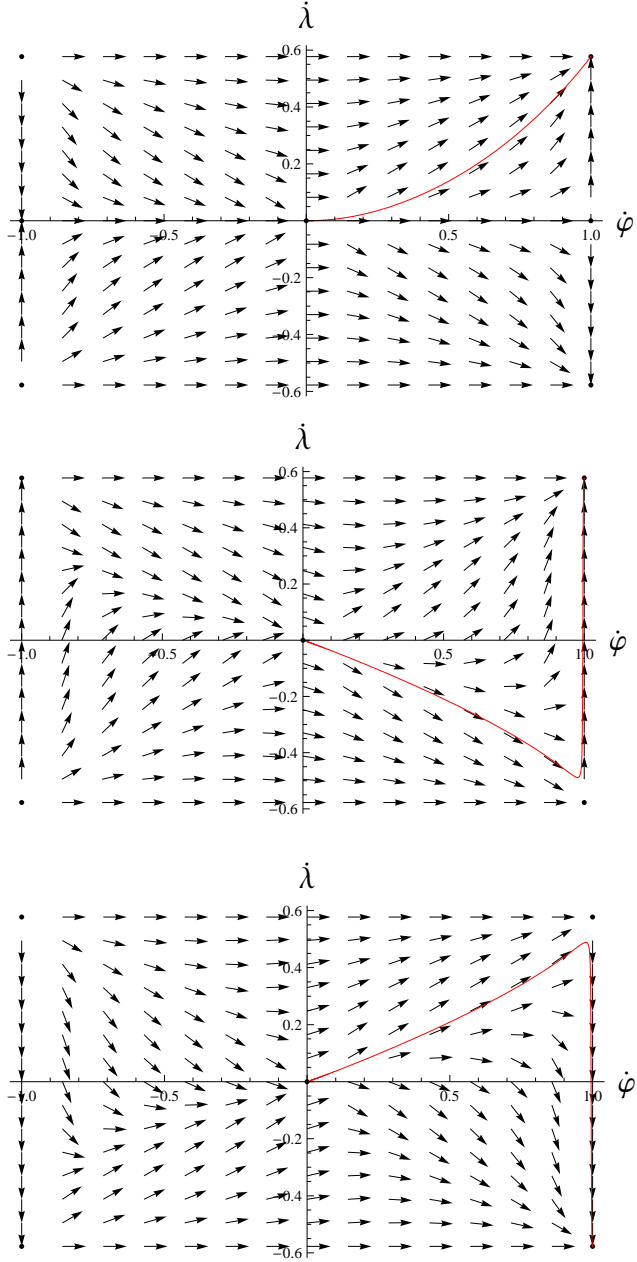}
\caption{{\small{Phase space flows for $w=0$ (top), $w = 1/d$ (middle), $w = -1/d$ (bottom).
 The five
 fixed points $(\dot{\phic},\dot{\lambda})=(\pm 1,\pm 1/\sqrt{d})$ and
 $(\dot{\phic},\dot{\lambda})=(0,0)$ are
 connected smoothly. Some typical trajectories are also shown. For $w=1/d$ and
 $w=-1/d$ they represent
 bounces of the scale factor due to KK and winding modes respectively.}}}
\label{flows}
\end{figure}

It is easy to see that these equations have fixed points at the
constant curvature, linear dilaton solutions
$(\dot{\phic},\dot{\lambda})=(\pm 1,\pm 1/\sqrt{d}$). These can be
smoothly connected to the trivial fixed point
$(\dot{\phic},\dot{\lambda})=(0,0)$ in the sense that no singularity
stands between them. This is an attractive feature that $\alpha'$
corrections to the low energy effective action are conjectured to
have, perhaps to all orders in $\alpha'$ \cite{gasp2}. It is
particularly relevant to pre-big bang models. The phase space flows
and some trajectories for $d = 3$ are shown in figure
\ref{flows}.\footnote{For $w \not= 0$ the equations of motion are
  singular when $\vert\dot{\phic}\vert = 1$.  This is not problematic
  because trajectories never quite reach points where
  $\vert\dot{\phic}\vert = 1$.  Instead the $\gl^{-1}\gp$ term in the
  equation for $\ddot{\lambda}$ eventually pushes the trajectories
  towards the line $\dot{\phic}=\pm\sqrt{d}\dot{\lambda}$ (depending
  on the sign of $w$) where $\gl^{-1}\gp\rightarrow 1$.}  In such a
smooth and ``connected'' phase space, the system can move around the
phase space towards the attractors without encountering singularities,
\textit{independently of initial conditions}.  This feature is hard to
obtain with generic $\alpha'$ corrections to dilaton gravity and is
crucial for the cyclic and bouncing solutions we will study below.
For example, with the conventional two-derivative action for dilaton
gravity one could at most hope for a single bounce before encountering
a singularity. In essence, with the new action, we have replaced these
singularities with the constant velocity fixed points. 

For general matter content there is a simple way to see how the
modified equations of motion capture the desired behavior.  Assuming
that $E$ is positive the Friedmann equation requires $\gp > \gl \geq
1$ so $\dot{\phic}$ can never vanish.  Orienting time so that
$\dot{\phic} < 0$, the dilaton rolls monotonically from strong to weak
coupling.  Since $\phic$ and $\lambda$ behave like relativistic
particles of mass $\sim e^{-\phic}$, at early times they are massless
and move at the speed of light:
\be
\label{relativistic}
\dot{\phic} \rightarrow -1 \qquad \dot{\lambda} \rightarrow \pm 1/\sqrt{d} \qquad\quad
\hbox{\rm as $t \rightarrow -\infty$.}
\ee
Thus at early times the scale factors grow exponentially and the
metric (\ref{metric}) approaches de Sitter space in planar
coordinates, with the spatial coordinates periodically identified to
make a torus.  This early-time de Sitter phase is what replaces the
big bang singularity in these models.\footnote{The dilaton diverges at
  $t = - \infty$, so strictly speaking we have not eliminated the
  singularity, just moved it infinitely far into the past.  As we will
  see even this can be cured by adding a potential for the dilaton.}
This is very reminiscent of the behavior obtained in
\cite{Mukhanov:1991zn}.  It is also similar to pre-big bang models
where the kinetic energy of the dilaton dominates and drives
inflation. One might worry about the fact that the coupling diverges
at early times; as we will see we can cure this behavior by
introducing a potential for the dilaton which violates positivity of
$E$.

\section{A first pass at thermodynamics\label{sect:thermo}}

To proceed further we need to specify the matter content of the
universe.  We will be fairly conservative at this stage, since we've
already modified the dilaton gravity action to eliminate
singularities.  Readers familiar with these standard results may skip
ahead to the next section.

We take matter to consist of the following ingredients.
\begin{enumerate}
\item
There may be a gas of string winding modes, characterized by winding
numbers $W_i$ that count the number of strings wound with positive
orientation around the $i{}^{th}$ dimension of the
torus.\footnote{Since we work in a compact space there must be an
equal number of strings wound with the opposite orientation.}  For
simplicity we set all $W_i = W$.  Then the energy in winding modes is
\[
E_W=2dWe^{\lambda}\,.
\]
\item
Likewise there may be a gas of Kaluza-Klein momentum modes,
characterized by positively-oriented momentum numbers $K_i$.  With
all $K_i = K$, the energy in Kaluza-Klein modes is
\[
E_K=2dKe^{-\lambda}\,.
\]
\item
We allow for a gas of string oscillator modes which we will
model as pressureless dust with energy $E_{\rm dust}$.
\end{enumerate}
To be precise, $W$ and $K$ refer to the winding and momentum numbers
in the first $d$ dimensions.  Thus we take $E_{\rm dust}$ to represent
the energy, not only in string oscillators, but also in winding and
momentum modes in the remaining $9-d$ dimensions.  These modes can be
modeled as dust since they do not contribute to the pressure in the
first $d$ dimensions.  As the remaining component of the energy budget,
we may introduce a potential for the dilaton $V(\phic)$.  The total
energy is then the sum
\begin{equation}\label{etot}
E=E_W+E_K+E_{\rm dust}+V\,.
\end{equation}
Treating the system adiabatically the ``pressures'' are
\bea
\label{Pdil}
&&P_\phic = {\partial E \over \partial \phic} = {\partial V \over \partial \phic} \\[4pt]
\label{pres}
&&P_\lambda = - {1 \over d} {\partial E \over \partial \lambda} = 2Ke^{-\lambda}-2We^{\lambda}
\eea
We will assume that the dilaton potential is independent of temperature.  However the other
components of the energy budget behave thermodynamically.  To make the
distinction, we refer to $E_s\equiv E_W+E_K+E_{\rm dust}$ as the
energy in the \textit{string gas}, the thermodynamical component of the total
energy. The following phases will be of interest to us \cite{Bassett:2003ck}.

\noindent
{\em Hagedorn phase} \\
In the Hagedorn phase we assume that all matter degrees of freedom are in thermal equilibrium at
the type II Hagedorn temperature $T_H = 1/(\sqrt{8}\pi)$.  Hagedorn thermodynamics has been studied
extensively \cite{brandvafa,deo0,deo1,bow0}.  The free energy of the string gas vanishes, so
$P_\lambda = 0$ and the energy $E_s$ is conserved; since $E_s = T_H S$ the entropy
is also conserved.  In equilibrium the winding and momentum numbers
are\footnote{These values follow from the distributions in \cite{deo1} with the assumption that the
energy is equally partitioned among all compact dimensions \cite{winds}.}
\begin{equation}
\langle W \rangle=\frac{\sqrt{E_s}}{12\sqrt{\pi}} \, e^{-\lambda} \hspace{1cm}
\langle K \rangle=\frac{\sqrt{E_s}}{12\sqrt{\pi}} \, e^{\lambda}
\end{equation}
As expected these values make the pressure in (\ref{pres}) vanish.

\noindent
{\em Radiation phase} \\
The radiation phase describes the equilibrium situation at temperatures $T < T_H$.  The universe is
dominated by a gas of massless string modes with energy
\be
\label{radE}
E_s = c_d V_d T^{d+1}\,.
\ee
Here $c_d$ is a constant appropriate to a gas of 128 massless Bose and 128 massless Fermi degrees of freedom,
\begin{equation}
c_d=128\frac{2d!\zeta(d+1)}{(4\pi)^{d/2}\Gamma(d/2)}(2-2^{-d})\,.
\end{equation}
Also $V_d = (2\pi)^d e^{d \, \vert \lambda \vert}$
is the T-duality invariant ``volume'' of the torus.  This definition takes into account the fact that the
energy could be stored in either momentum or winding modes depending on the size of the torus.  We have the
standard thermodynamic results
\bea
&&F = E_s - TS = - {1 \over d} c_d V_d T^{d+1} \\
&&P_\lambda = {\rm sign}(\lambda) E_s / d
\eea
leading as usual to a conserved entropy.  We also have the equilibrium values
\be
\label{RadiationValues}
\begin{array}{lll}
\lambda > 0: \qquad & \langle W \rangle = 0 \qquad & \langle K \rangle = {1 \over 2} P_\lambda e^{\lambda} \\[8pt]
\lambda < 0: \qquad & \langle W \rangle = - {1 \over 2} P_\lambda e^{-\lambda} \qquad & \langle K \rangle = 0
\end{array}
\ee
These follow from requiring that the pressure (\ref{pres}) takes on the correct value.

\noindent
{\em Frozen phase} \\
Finally as an alternative to an equilibrium radiation phase we
consider a frozen phase in which the interactions between strings are
turned off.  The momentum and winding numbers are conserved, so $K$
and $W$ are frozen at the values which they have on Hagedorn exit.
Any remaining energy in the universe goes into dust.  As we discuss in
section \ref{sect:entropy}, in this phase matter entropy is conserved.
In section \ref{sect:entropy} we will go beyond this approximation and
study entropy production due to interactions between winding and
momentum modes.  But please note that we will refer to an
out-of-equilibrium string gas as being in a radiation phase if the
equilibrium temperature would be below Hagedorn.

\section{Bouncing and cyclic cosmologies\label{sect:bouncing}}

In this section we study what happens when we couple the modified
dilaton-gravity action of section \ref{sect:modified} to a gas of
string winding and momentum modes.  For simplicity we will model the
string gas using just the Hagedorn and frozen phases described in
section \ref{sect:thermo}.  This is not very realistic, but it will
serve to illustrate the way a string gas changes the dynamics.  We
will give a more realistic treatment in section \ref{sect:exit}.

First let's see what happens for vanishing dilaton potential.  Whether
we're in a Hagedorn or frozen phase the matter energy is positive, so
as shown above (\ref{relativistic}) the dilaton will roll
monotonically from strong to weak coupling.  Suppose we're at strong
coupling, and let's assume we're in equilibrium in the Hagedorn phase
with $\dot{\lambda} > 0$.  Since we're at strong coupling the fields
$\phic$ and $\lambda$ behave like massless particles.\footnote{We will be
more precise about this in (\ref{fried2}) below.}
Moreover there's no force on these particles: with no dilaton
potential $P_\phic = 0$, and in the Hagedorn phase $P_\lambda = 0$.
So the particles move at nearly the speed of light,
\be
\label{relativistic2}
\dot{\phic} \approx -1 \qquad\quad \dot{\lambda} \approx 1/\sqrt{d}\,.
\ee
But this behavior cannot persist indefinitely.  As the universe
expands eventually it will cool below the Hagedorn temperature.  To
see when this happens we compute the energy $E$ in matter using the
Friedmann equation (\ref{encon}).  Then we compute the equilibrium
radiation temperature $T_{\rm rad}$ using (\ref{radE}).  If $T_{\rm rad} < T_H$
the universe is no longer in the Hagedorn phase.  But
rather than go to an equilibrium radiation phase, we assume the
universe makes a transition to a frozen phase in which the momentum
and winding numbers $K$ and $W$ are conserved, equal to whatever
values they had on Hagedorn exit.

\begin{figure}[htp]
\includegraphics[scale=0.9,viewport=25 580 520 720,clip]{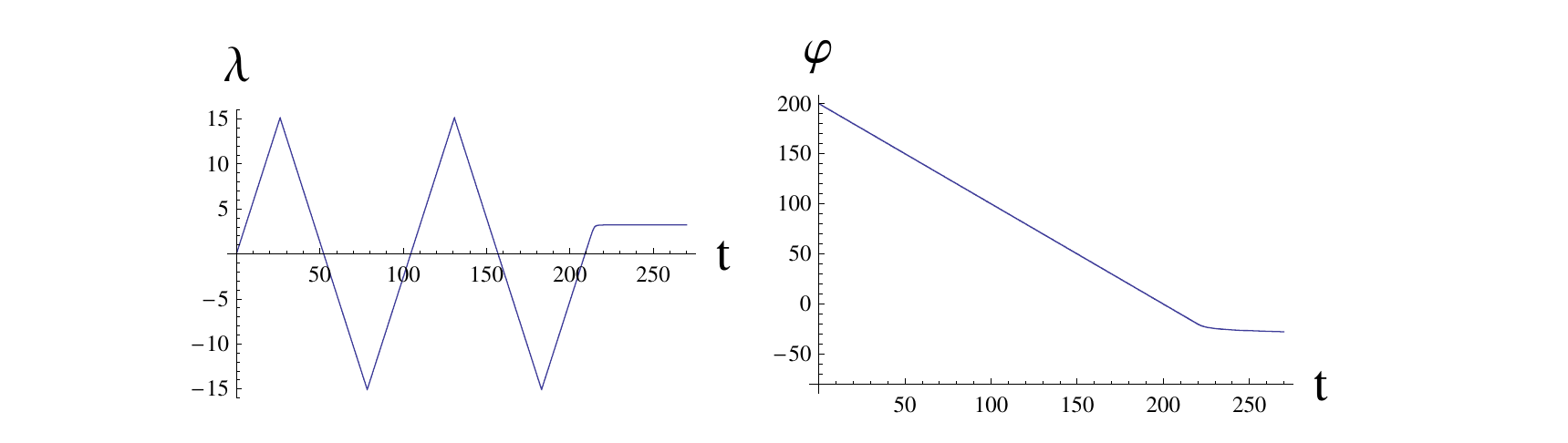}
\caption{{Numerical solution with Hagedorn and frozen phases
    and no potential for the dilaton. The oscillations have constant
    amplitude as there is no entropy production. The oscillations stop
    when the universe reaches weak coupling. We use $d=3$, as in all graphs
    that follow.}}
\label{bounc1}
\includegraphics[scale=0.85,viewport=20 560 520 720,clip]{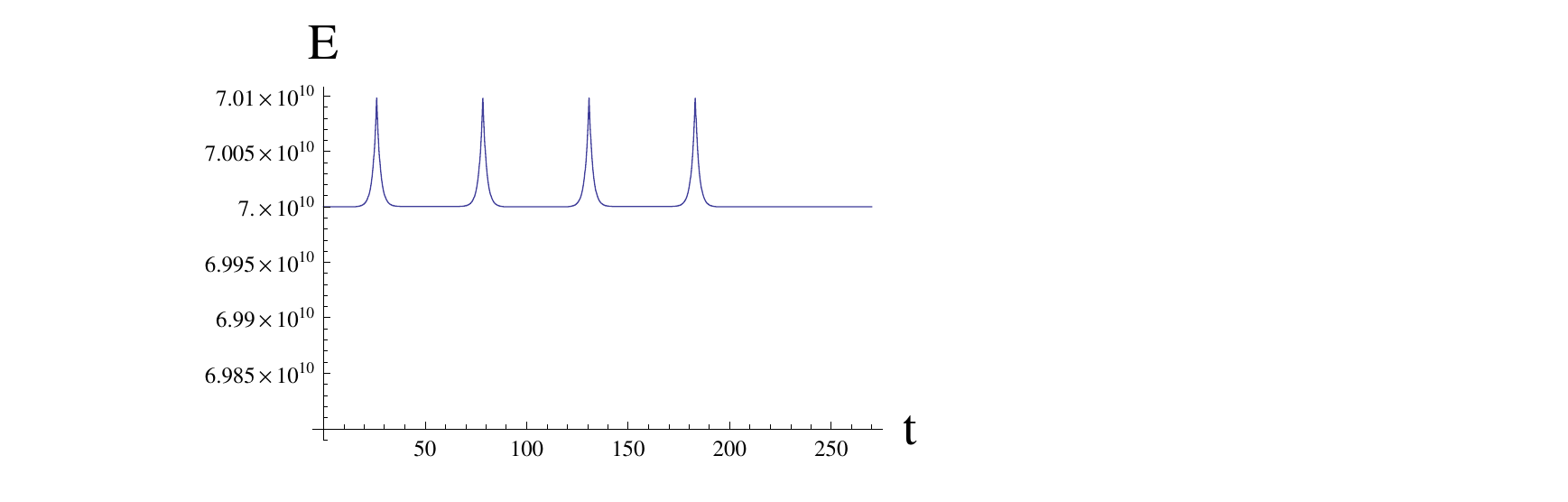}
\caption{{A plot of the energy in the string gas for Fig.~\ref{bounc1}. The
    energy is constant during the Hagedorn phases. During a frozen phase it
    increases until the scale factor bounces. It then decreases
    and the system re-enters the Hagedorn phase.}} 
\label{energy1}
\end{figure}

In the frozen phase the pressure does not vanish.  Instead there is an
effective potential for the scale factor,
\be
\label{LambdaPot}
V(\lambda) = E_W + E_K = 2dWe^\lambda + 2dKe^{-\lambda}\,.
\ee
At some point $\lambda$ bounces off this potential.  The universe
shrinks and eventually re-enters a Hagedorn phase.  It subsequently
emerges from this new Hagedorn phase and undergoes a T-dual bounce,
driven by momentum modes, at $\lambda < 0$.  The whole cycle repeats,
resulting in an oscillating scale factor.  However the oscillations
cannot persist indefinitely.  When the dilaton reaches weak coupling
the $\phi$ and $\lambda$ particles become very massive and come to
rest, putting an end to the oscillations.  This can be seen in a
numerical solution in figure \ref{bounc1}. Note that at strong
coupling the oscillations have constant amplitude.  This is a
consequence of neglecting interactions, which implies no entropy
production in the frozen phase: the system always re-enters the
Hagedorn phase with the same values of $\lambda$ and $E$, which in the
Hagedorn phase corresponds to the system having the same entropy.  We
will relax this approximation in section \ref{sect:entropy}.

What we need for a cyclic scale factor is not strong coupling, necessarily, but rather
a large amount of energy stored in the dilaton. This can be seen from the Friedmann equation
\begin{equation}\label{fried2}
\gl=\frac{1}{8\pi^2}(E_{\phic}-E_s)e^{\phic}
\end{equation}
where $E_{\phic}=8\pi^2\gp e^{-\phic}-V(\phic)$ is the (negative of) the total energy stored in the
dilaton, and $E_s$ is the energy in the string gas. As long as $E_{\phic}$ is large enough
the scale factor is relativistic and can undergo bounces in a suitable potential.

It is useful to note that when $\gl>>1$ the equations of motion
(\ref{eqmot}) imply $\dot{\gp} \approx \dot{\phic}\gp +
\frac{1}{8\pi^2}\dot{\phic}e^{\phic}\frac{\partial
V(\phic)}{\partial\phic}$, so $\frac{d}{dt}(8\pi^2\gp
e^{-\phic}-V(\phic))\approx 0$ and $E_{\phic}$ is conserved.  In the
Hagedorn phase $E_s$ is conserved as well, so (\ref{fried2}) gives a
clear picture of the dynamics: in the frozen phase, as $\lambda$ grows
the winding modes (or KK modes in the dual picture) absorb energy and
increase $E_s$ until $\gl$ drops to $1$ and the universe bounces.  A
plot of $E_s$ is shown in figure \ref{energy1}.

\begin{figure}[htp]
\includegraphics[scale=0.85,viewport=25 560 520 720,clip]{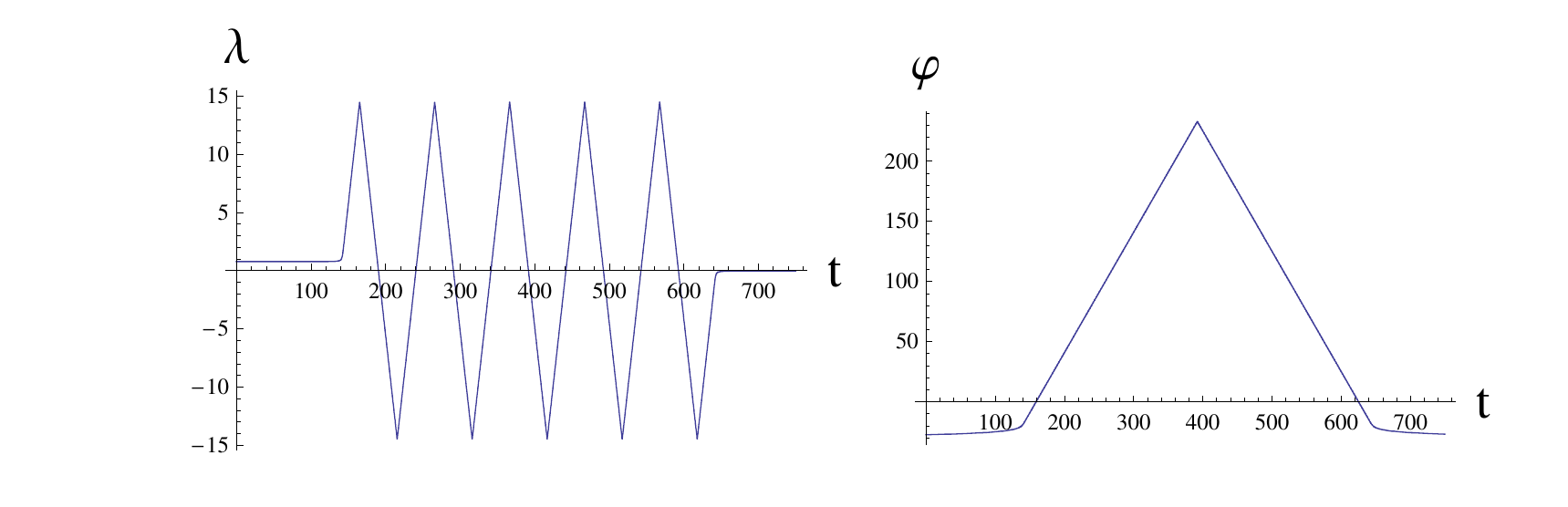}
\caption{{Same as Fig.~\ref{bounc1}, but with a dilaton potential of
    the form $Ae^{\phic}$ that yields a single bounce for the dilaton.}}
\label{likeprebb}
\includegraphics[scale=0.85,viewport=20 580 520 720,clip]{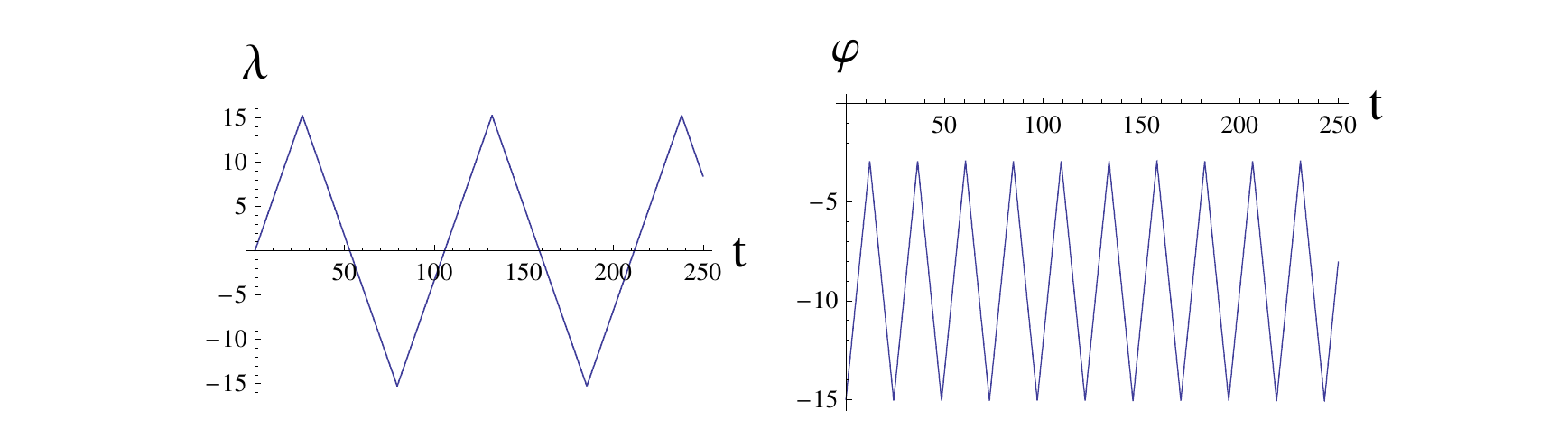}
\caption{{A potential of
    the form $Ae^{\phic}+Be^{-\phic}$ can confine the dilaton at weak coupling.}}
\label{dilpot2}
\end{figure}

So far we have discussed solutions in which the dilaton evolves
monotonically.  However the dilaton need not run to infinite coupling
in the far past. A past state for the universe could be one where the
expansion rate is arbitrarily small and the string coupling is
arbitrarily weak.  Provided $\dot{\phic}\gtrsim 0$ at early times, a
simple dilaton potential of the form $V(\phic)=Ae^{\phic}$ with $A<0$
can generate a bounce for $\phic$ and turn it back toward weak
coupling at late times. This is the basic idea of the pre-big bang
scenario.  A numerical solution is shown in figure \ref{likeprebb}.

As a further example, an upside down potential of the form $V(\phic) =
A e^{\phic} + B e^{-\phic}$, with $A$ and $B$ negative, can restrict
the dilaton to vary within a finite range. The dilaton will undergo
bounces, just like the scale factor, with $E_{\phic}$ converting
between large negative kinetic energy and large negative potential
energy.  A typical numerical solution is shown in figure
\ref{dilpot2}.\footnote{The initial conditions in figures \ref{bounc1},
\ref{likeprebb} and \ref{dilpot2} are chosen such that when
$\gl>>1$, $E_{\phic}$ has the same value in all three examples. The
Hagedorn phase $E_s$ is also chosen to be the same (smaller than
$E_{\phic}$). These two energies determine the amplitude of the
cycles, as we will see in more detail in section \ref{sect:exit}, so
the maximum value of $\lambda$ is the same in all three figures.}

So far we have discussed bouncing and cyclic behavior using the string
frame metric.  Since we have in mind coupling to stringy matter this is
the physically relevant frame to use.  However one might be interested in
the behavior of the Einstein frame metric, with scale factor
\[
\lambda_E = - {1 \over d - 1} (\varphi + \lambda)\,.
\]
If the matter energy is positive, implying that the dilaton evolves monotonically, then
the Einstein frame scale factor will evolve monotonically as well: the Friedmann
equation (\ref{encon}) requires $\dot{\phic}^2 > d \dot{\lambda}^2$.
However the models with a bouncing dilaton lead to a bouncing scale factor in
Einstein frame.  Generically each bounce of the dilaton will correspond to a bounce of $\lambda_E$.

One might worry that we have introduced dilaton potentials which are
unbounded below.  However note that our solutions only explore a
limited range of $\phic$, and one could easily imagine obtaining the
same behavior from a stable potential, just by modifying $V(\phi)$
outside the range of variation of the dilaton.  One might also worry
that bouncing and cyclic cosmologies require violation of certain
energy conditions.  We address this in appendix \ref{sect:Econd}.

\section{Interactions and entropy production\label{sect:entropy}}

In this section we study the effect of interactions on an out-of-equilibrium
string gas.  We will continue to assume that thermal equilibrium holds during
the Hagedorn phase, but we will allow the momentum and winding modes to go out
of equilibrium in the radiation phase, where the temperature is below Hagedorn.
We first take a macroscopic thermodynamic perspective and discuss entropy production,
then present Boltzmann equations for the winding and momentum numbers.  For simplicity
in this section we will neglect the possibility of having a dilaton potential.

Our goal is to understand how the momentum and winding numbers $K$ and $W$ evolve towards
their equilibrium values.  One constraint comes from energy conservation.  The equations of
motion (\ref{eqmot}) along with the Hamiltonian constraint (\ref{encon}) imply that
\begin{equation}
\label{pde}
\dot{E} = - d P_\lambda \dot{\lambda}\,.
\end{equation}
Here the dot indicates a time derivative and $d$ is the number of
dimensions, not a differential.  Breaking up the matter energy as in
(\ref{etot}), namely $E = E_W + E_K + E_{\rm dust}$, and likewise
breaking up the pressure, the energy conservation equation (\ref{pde})
becomes
\begin{equation}
\label{pde2}
\dot{E}_W+\dot{E}_K + \dot{E}_{\rm dust} = -d(P_W + P_K)\dot{\lambda}\,.
\end{equation}
For the individual species we have
\begin{equation}\label{ew}
\begin{aligned}
\dot{E}_W={d \over dt}(2dWe^{\lambda})&=2dWe^{\lambda}\dot{\lambda}+2d\dot{W}e^{\lambda}\\
&=-dP_W\dot{\lambda}+2d\dot{W}e^{\lambda}
\end{aligned}
\end{equation}
and
\begin{equation}\label{ek}
\begin{aligned}
\dot{E}_K={d \over dt}(2dKe^{-\lambda})&=-2dKe^{-\lambda}\dot{\lambda}+2d\dot{K}e^{-\lambda}\\
&=-dP_K\dot{\lambda}+2d\dot{K}e^{-\lambda}
\end{aligned}
\end{equation}
Combining (\ref{pde2}), (\ref{ew}) and (\ref{ek}), we must have
\begin{equation}\label{rates}
\dot{E}_{\rm dust}+2d(\dot{W}e^{\lambda}+\dot{K}e^{-\lambda})=0
\end{equation}
in order for energy to be conserved.

Another constraint comes from the second law of thermodynamics.  To
illustrate what's required let's temporarily model the universe as
filled with two fluids at different temperatures.  One fluid consists
of pressureless dust and winding modes and is held at the Hagedorn
temperature $T_H$, the other consists of radiation (i.e.\ momentum
modes) held at temperature $T_K$.  The two fluids are out of
equilibrium when $T_K < T_H$ which is what we expect to occur when we
exit the Hagedorn phase.  The resulting entropy production rate is
\begin{equation}
\begin{aligned}
\dot{S}&=\dot{S}_K+\dot{S}_W+\dot{S}_{\rm dust}\\
&=\frac{\dot{E}_K+dP_K\dot{\lambda}}{T_K}+\frac{\dot{E}_W+dP_W\dot{\lambda}}{T_H}
+\frac{\dot{E}_{\rm dust}}{T_H}\\
&=\frac{2d\dot{K}e^{-\lambda}}{T_K}+\frac{\dot{E}_{\rm dust}+2d\dot{W}e^{\lambda}}{T_H}
\end{aligned}
\end{equation}
and using (\ref{rates})
\begin{equation}
\dot{S}=2d\dot{K}e^{-\lambda}\left(\frac{1}{T_K}-\frac{1}{T_H}\right)\,.
\end{equation}
Provided $\dot{K}$ is positive (radiation is produced) whenever $T_K <
T_H$, we will have $\dot{S}>0$ consistent with the second law of
thermodynamics.  So in the frozen phase of section \ref{sect:thermo},
where $K$ was conserved, there was no entropy production.  But any
sensible evolution equation for $K$ and $W$ will lead to an increase
in entropy.

We now present such an evolution equation for the winding number $W$.
At weak string coupling the appropriate Boltzmann equation was derived
in \cite{winds}, based on the cross section for winding --
anti-winding annihilation obtained in \cite{polch}.
\begin{equation}
\label{bolt1weak}
\dot{W}=-\frac{e^{2\lambda + \phic}}{\pi}\left(W^2 - \langle W \rangle^2\right)
\end{equation}
Here $\langle \, \cdot \, \rangle$ denotes a thermal expectation
value, given in (\ref{RadiationValues}) for temperatures below
Hagedorn.  This expression makes intuitive sense: the factor
$e^{2\lambda}$ captures the fact that longer strings are more likely
to annihilate, while the factor $e^\phic = g_s^2 / V$ takes into
account both enhancement by the string coupling $g_s$ and suppression
by the volume of the torus $V$.  The result (\ref{bolt1weak}) is reliable
at weak coupling, but we will often be interested in behavior at
strong coupling.  At strong coupling we adopt the following modified Boltzmann
equation.
\begin{equation}\label{bolt1}
\dot{W}=-\frac{e^{2\lambda-d|\lambda|}}{\pi}\left(W^2 - \langle W \rangle^2\right)
\end{equation}
This equation can be obtained from the
previous weak-coupling Boltzmann equation (\ref{bolt1weak}) by making
the replacement $\phic \rightarrow -d|\lambda|$, that is, by dropping
the unshifted dilaton from the cross section but keeping the
dependence on the T-duality-invariant ``volume'' $\exp(-d \vert
\lambda \vert)$.  This can be thought of purely phenomenologically, as
describing winding strings (such as cosmic strings) whose interactions do not
depend on the unshifted dilaton.  It can also be regarded as describing fundamental
strings, but with a potential for the dilaton that fixes the unshifted dilaton to
$\phi \approx 0$.  For momentum modes at strong coupling we use the
T-dual equation
\begin{equation}\label{bolt2}
\dot{K}=-\frac{e^{-2\lambda-d|\lambda|}}{\pi}\left(K^2-\langle K\rangle^2\right)\,.
\end{equation}

\section{Shrinking cycles and exit\label{sect:exit}}

We now study how entropy production in an out-of-equilibrium string
gas affects the cyclic cosmologies of section \ref{sect:bouncing}.
For simplicity we set the dilaton potential to zero.

Recall that in section \ref{sect:bouncing} we neglected interactions
during the frozen phase; the momentum and winding numbers were taken
to be conserved.  This led to a constant entropy and oscillations of
fixed amplitude.  Taking interactions into account we will see that
the resulting entropy production leads to oscillations of decreasing
amplitude.  Oscillating models often exhibit this sort of behaviour, but the
details depend on the mechanism that drives the bounce \cite{clifton}.
For example in a recent bouncing cosmology, in which an equilibrium
Hagedorn era was also used, the oscillations grew with time
\cite{biswas}.  But in this model there was no dilaton and the bounce
was driven by positive spatial curvature and negative Casimir energy.

In our models eventually so much entropy is produced that it is no
longer thermodynamically possible for
the universe to re-enter the Hagedorn phase.  At this point the
universe transitions to a loitering phase in which the scale factors
are roughly constant, oscillating about a minimum in their potential.
Eventually the loitering phase also ends and the universe transitions
to a standard radiation-dominated cosmology.

\subsection{Shrinking cycles}

The dynamics are largely governed by the energy stored in the
dilaton. We are neglecting any dilaton potential, so as noted
in section \ref{sect:bouncing} the (negative of) the dilaton
kinetic energy
\be
\label{Emax}
E_{\rm max} \equiv 8 \pi^2 e^{-\phic} \gamma_\phic
\ee
is essentially constant.  We have denoted this $E_{\rm max}$ because it's
equal to the maximum matter energy during a cycle.  To see this recall
that the Friedmann equation (\ref{encon}) states that the energy in
matter is
\be
\label{shrink:encon}
E = 8 \pi^2 e^{-\phic} \left(\gp - \gl\right)\,.
\ee
At a bounce we have $\gp \gg \gl = 1$ and therefore $E \approx E_{\rm max}$.

During the radiation phase of the $n^{th}$ cycle the energy in matter
starts at $E_n$, the (conserved) matter energy during the Hagedorn
phase of the $n^{th}$ cycle.  It increases to $E_{\rm max}$ as the
wound strings are stretched.\footnote{For simplicity we discuss
bounces at large radius.  At small radius T-duality would exchange
momentum and winding.}  After the bounce the matter energy decreases
down to the value $E_{n+1}$ associated with the next Hagedorn phase.
These Hagedorn phases serve as reference equilibrium points in phase
space where the entropy is well defined, given by $S_n=E_n/T_H$.
Since entropy is produced during the radiation phase, $S_{n+1} > S_n$
as we saw above, and since we return to the same (equilibrium)
temperature $T_H$ when re-entering the Hagedorn phase, the matter
energy must increase during each radiation phase as well, $E_{n+1} >
E_n$.  This means the radius at which we exit the Hagedorn phase also
increases with each cycle.  To see this recall that the condition for
exit is that the equilibrium radiation temperature drops below
Hagedorn.\footnote{Note that no \emph{real} temperature is dropping
  here since during the Hagedorn phase the temperature is constant at $T_H$. By equilibrium radiation temperature we mean the
  temperature that radiation alone would have in a universe of
  volume $V=(2\pi)^de^{d\lambda}$ and energy $E_n$. It is the volume that
  grows and signals
a transition to a radiation phase.}  From (\ref{radE}) this means that at Hagedorn exit
\be
\label{En}
E_n = c_d \, (2\pi)^d e^{d\lambda_n} \, T_H^{d+1}\,.
\ee
Since $E_n$ increases with each cycle, so does the scale factor at
exit $e^{\lambda_n}$.

We can also estimate the maximum scale factor reached during each cycle
$e^{\lambda_n^{\rm max}}$.  From the moment of Hagedorn exit to the bounce,
matter energy increases by an amount
\begin{equation}
E_{\text{max}} - E_n = -d \int_{\lambda_n}^{\lambda_n^{\text{max}}} d\lambda\,
P_\lambda \approx 2d \int_{\lambda_n}^{\lambda_n^{\text{max}}} d\lambda \,
\left(W_n e^{\lambda} - K_n e^{-\lambda}\right)
\end{equation}
where we've assumed interactions are weak so the values at Hagedorn exit
\be
\label{ExitValues}
W_n={\sqrt{E_n} \over 12 \sqrt{\pi}}e^{-\lambda_n} \qquad\quad
K_n={\sqrt{E_n} \over 12 \sqrt{\pi}}e^{\lambda_n}
\ee
are roughly conserved.  This leads to
\begin{equation}
e^{\lambda_n^{\text{max}}} \sim E_n^{1/d} \left(\alpha(E_n)+\sqrt{\alpha(E_n)^2 - 1}\right)
\end{equation}
where
\[
\alpha(E_n) = \frac{3}{d}\sqrt{\frac{\pi}{E_n}}\left(E_{\text{max}}-E_n\right)+1\,.
\]
$\lambda_n^{\rm max}$ is a decreasing function of $E_n$, so the maximum
radius shrinks with each cycle.

\begin{figure}[htp]
\includegraphics[scale=0.85,viewport=80 565 600 750,clip]{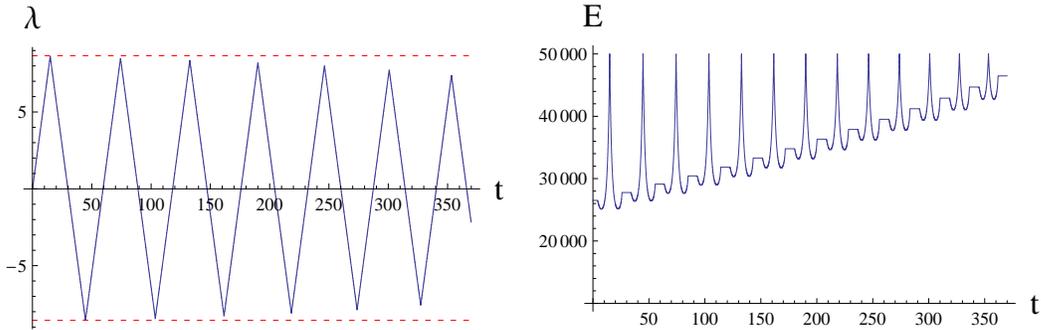}
\caption{{An integration of the equations of motion
    (\ref{eqmot}), (\ref{encon}), (\ref{bolt1}), (\ref{bolt2}) for
    $d=3$. As the entropy increases the energy during the Hagedorn
    phases increases towards $E_{\rm max}$ and the size of the
    oscillations in the scale factor gets smaller (the dashed lines
    are drawn at constant $\lambda$).  }}
\label{sh1}
\end{figure}

The features we have discussed can be seen in figure \ref{sh1}, which
shows a numerical solution to the combined gravitational equations of
motion (\ref{eqmot}), (\ref{encon}) and the strong-coupling Boltzmann
equations (\ref{bolt1}), (\ref{bolt2}).  The matter energy $E$ has
plateaus which correspond to Hagedorn phases of vanishing pressure.
During the radiation phases the matter energy jumps to $E_{\rm max}$
before falling to the next Hagedorn plateau.\footnote{The small dips
in the energy on either side of the plateaus is due to the redshift
of energy in an expanding radiation-dominated universe if $\lambda >
0$, or the T-dual phenomenon if $\lambda < 0$.  Eventually either
the stretching of winding strings its T-dual takes over and leads to
the large spikes in energy.  Note that time-reversal invariance is
only broken by entropy production during the radiation phases.}  In
figure \ref{sh1} one can also see the slight decrease in the amplitude
of the oscillations with time.\footnote{In figure
  \ref{sh1} we used by hand a slightly larger value for $T_H$ (larger by a factor
  of 1.7). This allows us to illustrate the desired effects over a
  shorter integration time as the phase transitions between Hagedorn and
  radiation phases occur earlier (smaller $\lambda$) and the interactions are
  more efficient. The qualitative picture is not altered.}

\subsection{After the Hagedorn era}

The dilaton kinetic energy $E_{\text{max}}$ sets the maximum possible
entropy that the system can have and still be in the Hagedorn phase,
namely $S_{\text{max}}=E_{\text{max}}/T_H$.  As entropy is produced
during the radiation phases eventually a bounce will occur during
which $S$ exceeds $S_{\rm max}$.  At this point a return to a Hagedorn
phase is no longer possible.

Instead the universe enters a new era which resembles the loitering
phase discussed in \cite{Brandenberger:2001kj}.  The scale factor
undergoes oscillations about the minimum of the potential
(\ref{LambdaPot}), namely\footnote{As can be seen from
(\ref{ExitValues}), at the moment of Hagedorn exit $W$ and $K$ are
such that the scale factor sits at the minimum of the potential.
For $\lambda > 0$ the subsequent evolution of $W$ and $K$ will tend
to shift the minimum to larger radii.}
\[
V(\lambda) = 2dWe^\lambda + 2dKe^{-\lambda}\,.
\]
Assuming $\lambda > 0$, and using the strong-coupling Boltzmann
equations (\ref{bolt1}), (\ref{bolt2}), the winding strings will
gradually annihilate and radiation (momentum modes) will be
produced.\footnote{With the weak-coupling Boltzmann equation
(\ref{bolt1weak}) the strings may never annihilate since the
interaction rate turns off as the dilaton rolls to weak coupling
\cite{winds,danos}.}  Eventually all the winding strings will be
gone.  At that point the oscillations stop and the universe
transitions to a radiation-dominated cosmology.  With our modified
gravity action we may not have the usual radiation-dominated
expansion, as one could enter the radiation-dominated era while the
scale factors and shifted dilaton are still relativistic ($\dot{\phic}
\approx -1$, $\dot{\lambda} \approx 1/\sqrt{d}$).  But eventually the
matter energy, or better the combination $E e^\phic$, becomes small
enough that the higher-derivative modifications to the action are
unimportant and we go over to a standard radiation-dominated
cosmology.  The dilaton continues to roll to weak coupling, while the
scale factor grows according to
\[
e^\phic \sim {1 \over t^{2d/(d+1)}} \qquad\quad e^\lambda \sim t^{2/(d+1)}\,.
\]
Somewhat curiously the unshifted dilaton is constant and the scale
factor grows just as it would in Einstein gravity.

\begin{figure}[htp]
\begin{center}
\includegraphics[height=7.8cm]{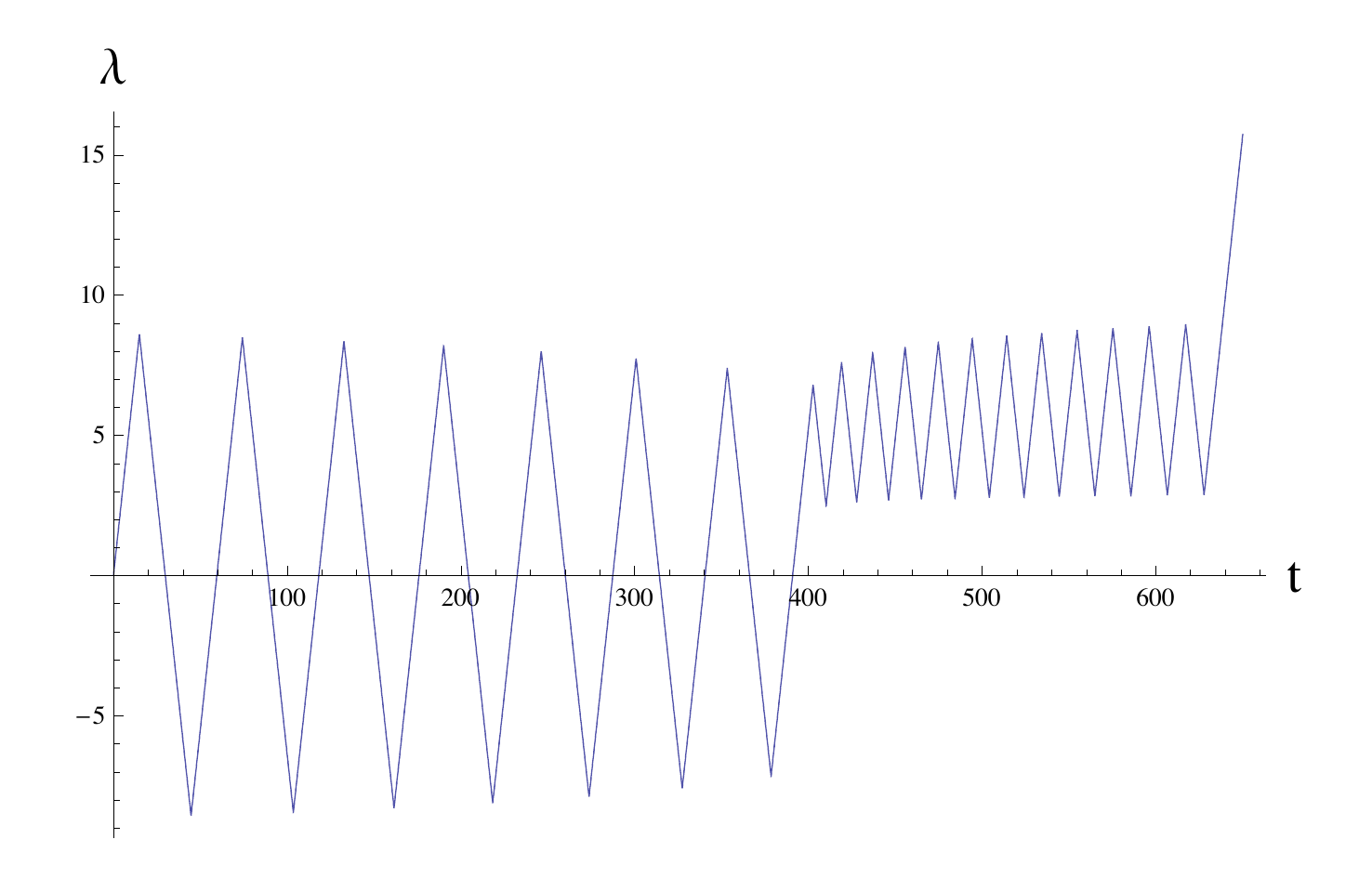}
\includegraphics[height=7.8cm]{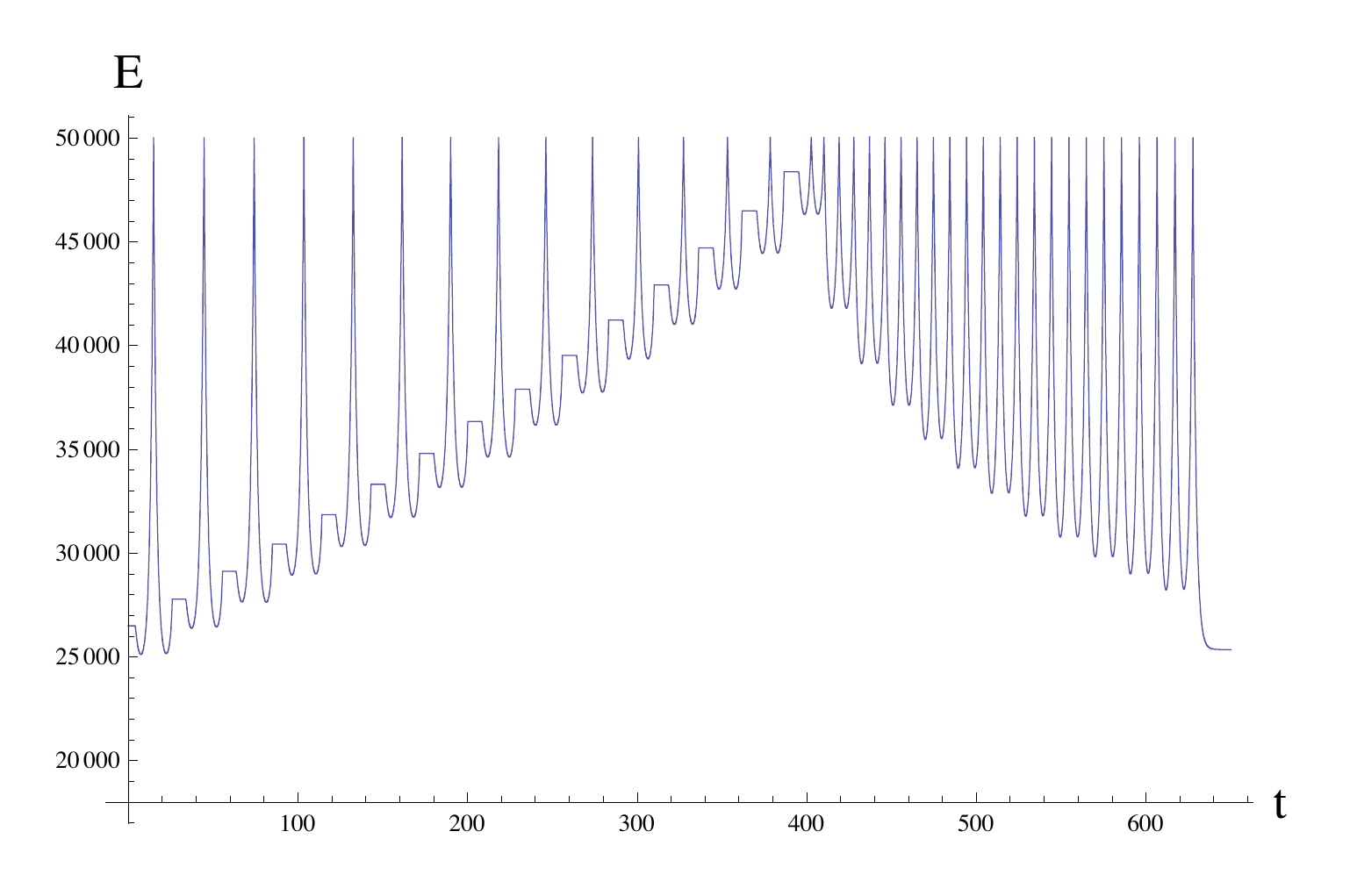}
\end{center}
\caption{{The log of the scale factor and the matter energy in
    a typical numerical solution.  For $t < 400$ the universe cycles
    between Hagedorn and radiation phases.  For $400 < t < 640$ the
    scale factor oscillates about the minimum of its potential while
    the winding strings gradually annihilate (in practice we use a cutoff
    value $W=1/2$ to specify winding mode annihilation).  For $t > 640$ the
    universe is radiation-dominated.}}
\label{exit}
\end{figure}

The whole story can be seen in figure \ref{exit} which is simply an
extension of figure \ref{sh1} to later times.  It shows the log of the
scale factor and the matter energy for a universe evolving through an
era of Hagedorn oscillations and a loitering era of potential
oscillations before finally entering a radiation-dominated era.  In
the Hagedorn era the scale factor oscillates about $\lambda = 0$,
while in the loitering era it oscillates about the minimum in the
potential, and in the radiation-dominated era it starts out growing
relativistically.  The amplitude of the oscillations decreases during
the Hagedorn era and increases during the loitering era.  The behavior
of the matter energy also changes.  It has plateaus during the era of
Hagedorn oscillations which disappear during the loitering era.  (The
spikes in the matter energy during the loitering era are simply
conversion between kinetic and potential energy.)

\section{On the BV decompactification mechanism\label{sect:BV}}

One might expect the models we have been discussing to provide an
ideal setting for realizing the Brandenberger-Vafa mechanism.  Indeed
this was our original motivation for developing these models.  The
original BV scenario runs into two difficulties \cite{winds,danos}: as
the dilaton rolls to weak coupling, the standard Boltzmann equation
(\ref{bolt1weak}) predicts that string interactions turn off, and one
is generically left with a gas of non-interacting strings on a torus
of fixed size.  Also with the two-derivative effective action
(\ref{s0}) the universe has a singularity a finite time in the past,
so there is only a limited amount of time for the necessary thermal
fluctuations to take place.

Both of these difficulties would seem to be cured in the models we
have considered.  With the modified Boltzmann equations (\ref{bolt1}),
(\ref{bolt2}) string interactions do not turn off at late
times.\footnote{This could also be achieved with the standard
Boltzmann equations by introducing a potential to confine the
dilaton.}  Moreover with the modified gravity action (\ref{s}) the
singularity is pushed infinitely far into the past.  The oscillating
scale factors we have found can be thought of as repeated attempts at
decompactifying; if on each bounce there was some probability of
decompactifying for $d \leq 3$, but vanishing probability for $d \geq
4$, then the Brandenberger-Vafa mechanism would work.

This is not, however, the behavior we generically find.  Instead in
any number of dimensions the era of Hagedorn oscillations eventually
ends and the universe transitions to a loitering phase of oscillations
about the minimum of the effective potential for $\lambda$.  Taking
$\lambda > 0$ for purposes of discussion, with the modified Boltzmann
equation the winding strings will eventually annihilate and the
universe will decompactify.  This chain of events can happen for any
$d$.  In this sense the Brandenberger-Vafa mechanism is not operative.

The reason why the BV mechanism seems to be failing is that quantum
fluctuations give the winding strings an effective thickness of order
$\sqrt{\alpha'}$ in all spatial dimensions, hence their probability to
interact is non-vanishing for any $d$ and only decreases with $d$
through a ``per volume,'' $\sim e^{-d\lambda}$, dependence. The
original BV argument rested on interactions via (classical) string
intersections which did not take into account this (quantum)
thickness.

One might still hope that $d \leq 3$ is favored because the universe
might not follow the expected behavior we discussed above.  Imagine
that due to a thermal fluctuation the universe exits a Hagedorn phase
with an unusually small number of winding strings.  In the subsequent
radiation phase perhaps all these strings will annihilate and the
universe will decompactify immediately, without additional bounces and
without going through a loitering era.  Since the annihilation rate
(\ref{bolt1}) falls off rapidly with $d$, perhaps this
fluctuation-driven mechanism will preferentially decompactify $d \leq
3$?

To address this issue let's study the conditions for decompactifying
in a single cycle in the framework we have been using.  The
probability of decompactifying depends not only on the energy during
the Hagedorn phase $E_n$, which determines the number of winding
strings present at Hagedorn exit, but also on $E_{\rm max}$, which
determines how long the subsequent radiation phase will last.  For
fixed $E_{\rm max}$, smaller values of $E_n$ -- that is, less winding
on Hagedorn exit and a larger value of $\dot{\lambda}$ -- will give an
increased probability of decompactifying.  It's convenient to express
this in terms of
\[
c = (E_{\rm max} - E_n) / E_{\rm max}\,.
\]
Since $E_n > 0$ we have $c < 1$.  Taking $E_{\text{max}} = 10^7$ as an
example, we find that 3 dimensions decompactify promptly on Hagedorn
exit for $c \gtrsim 0.984$.  To decompactify 4 dimensions requires $c
\gtrsim 0.9994$, and to decompactify more dimensions requires slightly
larger $c$.  To translate this into winding numbers on Hagedorn exit
we use (\ref{En}), (\ref{ExitValues}).  We find that 3 dimensions
decompactify promptly if $\langle W_n \rangle < 0.507$, while 4
dimensions decompactify promptly if $\langle W_n \rangle < 0.504$.  We
conclude that, with this value of $E_{\rm max}$, strings are only
slightly more efficient at annihilating in $d=3$ compared to $d=4$.
The only way to decompactify promptly is to exit Hagedorn with
essentially no winding (recall that our criterion for no winding was
$W < 0.5$).

One could imagine choosing special initial conditions -- say a small
value of $E_n$ -- to make the winding number small.  But this seems
against the spirit of the BV mechanism, which should operate starting
from generic initial conditions.  To quantify just how special the
initial conditions have to be, note that the number of Hagedorn-era
microstates which decompactify promptly (proportional to the
probability of decompactifying) is
\[
e^{S_n} = e^{E_n/T_H} \sim e^{-c E_{\text{max}} / T_H}\,.
\]
With $E_{\text{max}} = 10^7$ the probabilities of having sufficiently small
$E_n$ are tiny, although they do fall off rapidly with $d$.

Other types of fluctuations are more likely.  For example, even for
large $E_n$ and $\langle W_n \rangle$, there might be fluctuations
away from the mean that make the winding number vanish.  To
estimate the probability of this happening, note that for reasonable
distributions of winding numbers the probability of having zero
winding on Hagedorn exit scales as
\[
{\rm Prob.}(\hbox{\rm no winding}) \sim \big(1/\langle W_n \rangle\big)^d.
\]
If $\langle W_n \rangle$ is large then the probabilities are
tiny (although again they fall off rapidly with $d$).

There are other interesting types of fluctuations to consider, for
example fluctuations in the initial value of $\lambda$.  For large
initial $\lambda$ strings should be more likely to annihilate in $d=3$
than $d=4$, due to the dimension dependence of the cross-section.
Although we have not estimated the probability of this happening, it
seems unlikely to us that the basic picture will be modified:
fluctuations which are large enough to favor $d=3$ are also very
unlikely to take place. As an alternative approach, one could set
initial conditions such that three dimensions decompactify, but this
requires careful tuning and violates the spirit of the BV scenario.

We conclude with a few comments on the robustness of our results.  We
have assumed that the string gas is in thermodynamic equilibrium
during the Hagedorn phase.  Let's consider the alternative possibility
that the winding modes fall out of equilibrium during the Hagedorn
phase, well before exit to the radiation phase.  To test this we
should modify the Boltzmann equations to take into account the fact
that string oscillators are highly excited.  This was considered in
\cite{winds} and it amounts to putting a factor of $E$ in the string
cross-sections.  With this enhancement, numerical tests for a wide
range of energies ($10^3$ -- $10^7$) showed that during the Hagedorn
phase the winding number indeed closely tracked its thermodynamic
average.  This supports our assumption of a string gas in thermal
equilibrium during the Hagedorn phase.\footnote{Note however that in
  our models the collective degrees of freedom $\lambda$, $\phi$
  remain out of equilibrium with the rest of the string gas.  Showing
  that this is realistic, and not say an artifact of our truncation to
  homogeneous field configurations, deserves further study.  We are
  grateful to Matt Kleban for raising this issue.}  Another area of
concern is that we have modeled the Hagedorn $\rightarrow$ radiation
transition rather crudely (the average winding number jumps abruptly
from $\sim \sqrt{E}$ to $\sim E$).  It should be possible to do a
better job with the transition, using results of \cite{deo0}, but we
see no reason that an improved treatment of the transition should
favor $d = 3$.

\section{Conclusions\label{sect:conclusions}}

To summarize, we studied the dynamics of a string gas coupled to a
modified gravity action.  The modified gravity action was set up to
avoid singularities, and when coupled to a string gas we found that
bouncing and cyclic cosmologies naturally result.  Several aspects of our
analysis deserve comment and further investigation.
\begin{itemize}
\item
We postulated a particular form for the modified gravity action
(\ref{s}).  It would be interesting to understand to what extent our
action captures the effect of $\alpha'$ corrections in string
theory.  But we expect that any action which avoids singularities
and respects T-duality should lead to qualitatively similar results.
\item
Modified gravity theories generically have ghosts \cite{Chiba:2005nz}.  A crucial
question for future investigation is whether our action for the scale factors (\ref{s}) can be
lifted to a covariant theory, along the lines of \cite{Brandenberger:1998zs,Mukhanov:1991zn},
and whether the resulting theory is ghost-free \cite{InProgress}.
\item
We described the string gas using modified Boltzmann equations
(\ref{bolt1}), (\ref{bolt2}) in which we simply dropped the
dependence on the (unshifted) dilaton.  This could be thought of
quasi-phenomenologically, as describing cosmic strings whose
interactions do not depend on the dilaton.  It
could also be thought of as a crude representation of the behavior
of either fundamental strings or D-strings \cite{JJP}, given a potential
which confines the dilaton to string couplings $g_s = {\cal O}(1)$.
\item
Although we developed our models to illustrate some of the features
that result from a non-singular string gas cosmology, it would be
interesting to study whether they provide a basis for a realistic
cosmology.  An important step would be to study the spectrum of scalar perturbations
resulting from early Hagedorn and loitering eras, extending the work of
\cite{Nayeri:2005ck,Brandenberger:2006vv,Kaloper:2006xw} to the present context.
\end{itemize}
The models we have discussed provide a remarkably simple realization
of bouncing and cyclic cosmologies.  With a suitable potential for the
dilaton, they also provide a simple realization of the pre-big bang
scenario.  Let us comment on the two other motivations given in the
introduction.

\noindent
{\em Horizon problem} \\
As we have seen the universe can evolve to a loitering phase in
which the scale factor oscillates about the minimum of its potential.
If the loitering phase lasts long enough the entire universe will be
in causal contact and might be expected to become quite homogeneous.
This would provide a solution to the horizon problem.  There are two
conditions that must be met.
\begin{enumerate}
\item
The time-averaged scale factor $e^{\bar{\lambda}}$ depends on initial
conditions while the duration of the loitering phase $t$ also depends on the
string cross-section.  The condition for the universe to come in causal contact
is $e^{\bar{\lambda}} \ll t$ which can easily be satisfied by going to weak coupling.
\item
Even if the universe is in causal contact we still need to make sure it becomes homogeneous.
The condition is that the universe be smaller than the Jeans length, $e^{\bar{\lambda}}
\ll 1/\sqrt{G\rho}$.  Again this can easily be satisfied by going to weak coupling.\footnote{The
Jeans length during the Hagedorn phase was studied in \cite{Lashkari:2008sv} and argued to be small.  Here we
are interested in the Jeans length during the loitering phase for which we adopt
the naive estimate $1/\sqrt{G\rho}$.}
\end{enumerate}
Provided these conditions are satisfied any inhomogeneities generated
during the Hagedorn phase transitions will be washed out and the universe
will eventually approach radiation domination in a state very near thermal
equilibrium.\footnote{For a study of perturbations in bouncing models see
\cite{Martin:2003bp}.}  In this way our models provide a simple natural
resolution of the horizon problem.

\noindent
{\em Brandenberger-Vafa mechanism} \\
The Brandenberger-Vafa mechanism is predicated on the idea that
winding strings can only annihilate efficiently in $d \leq 3$
dimensions.  Our models evade this reasoning because the universe is
expected to enter a loitering phase in which we have strings with a
fixed coupling wound on a torus of fixed average size.\footnote{The
fixed coupling is due to our use of the modified Boltzmann equations
(\ref{bolt1}), (\ref{bolt2}).  With the standard Boltzmann equation
(\ref{bolt1weak}) the strings generically become non-interacting
before they can annihilate \cite{winds,danos}.}  These wound strings
will inevitably annihilate, and the universe will transition to
radiation domination, no matter the number of dimensions.  In this way
the Brandenberger-Vafa mechanism is not operative in the models we
have constructed.  We leave it as an open challenge to construct a
string-inspired model which does preferentially decompactify three
dimensions.

\bigskip
\goodbreak
\centerline{\bf Acknowledgements}
\noindent
We are grateful to Robert Brandenberger, Richard Easther, Mark
Jackson and Matt Kleban for numerous discussions on these matters.  BG and DK are
supported by DOE grant DE-FG02-92ER40699.  SM is supported by the Cyprus State
Scholarship Foundation.

\appendix
\section{Energy conditions in dilaton gravity\label{sect:Econd}}

Within Einstein gravity a bouncing cosmology requires $\rho + p < 0$, a violation of the
null energy condition \cite{Novello:2008ra,MolinaParis:1998tx}.  Here we make the
analogous statements for dilaton gravity.  A more detailed discussion can be found in
\cite{Kaloper:2007pw}.

With conventional (two-derivative) dilaton gravity the Friedmann equation and the
equations of motion are
\bea
\label{Econd1}
&& {1 \over 2} \dot{\phic}^2 = {1 \over 2} d \dot{\lambda}^2 + {1 \over 8 \pi^2} e^\phic E \\
\label{Econd2}
&& \ddot{\phic} = {1 \over 2} \dot{\phic}^2 + {1 \over 2} d \dot{\lambda}^2 + {1 \over 8 \pi^2}
e^\phic P_\phic \\
\label{Econd3}
&& \ddot{\lambda} = \dot{\phic} \dot{\lambda} + {1 \over 8 \pi^2} e^\phic P_\lambda
\eea
To study a bounce in the scale factor we set $\dot{\lambda} = 0$.
Then (\ref{Econd1}) requires $E \geq 0$.  Equation (\ref{Econd2})
gives no constraint, while (\ref{Econd3}) implies that $P_\lambda$ has
the same sign as $\ddot{\lambda}$.  A string gas can exert pressure of
either sign, so it is easy to obtain a bouncing or cyclic scale factor
in dilaton gravity coupled to a string gas.

To obtain a bounce in the dilaton is more difficult.  Setting
$\dot{\phic} = 0$ note that (\ref{Econd1}) requires $E \leq 0$.
Indeed we had to introduce negative potentials in section
\ref{sect:bouncing} to make the dilaton bounce.  Equation
(\ref{Econd3}) gives no constraint, while (\ref{Econd2}) can be
rewritten as
\be
\label{Econd4}
\ddot{\phic} = - {1 \over 8\pi^2} e^\phic (E - P_\phic)\,.
\ee
The sign of $\ddot{\phic}$ is correlated with the sign of $E - P_\phic$.  In particular
$\ddot{\phic} > 0$ requires $E - P_\phic < 0$, the dilaton gravity analog of violating
the null energy condition.\footnote{Due to the wrong-sign kinetic term for the dilaton we
inserted a minus sign in our definition of $P_\phic$ below (\ref{eqmot}).}

One can likewise study the conditions for a bounce in the Einstein-frame scale factor
$\lambda_E = - (\varphi + \lambda)/(d-1)$.  When $\lambda_E$ bounces we have $\dot{\varphi} =
- \dot{\lambda}$ and (\ref{Econd1}) requires $E < 0$.  Adding (\ref{Econd2}) and (\ref{Econd3})
gives
\[
\ddot{\lambda}_E = {d - 1 \over 8 \pi^2} e^\varphi (E - P_\varphi - P_\lambda)\,.
\]
Thus the sign of $\ddot{\lambda}_E$ is correlated with the sign of $E - P_\varphi - P_\lambda$.

The use of our higher-derivative modified action for dilaton gravity does not significantly
change these string-frame results.  In fact the only change is that (\ref{Econd4}) is replaced with
\[
\gamma_\lambda \ddot{\phic} = - {1 \over 8 \pi^2} e^\phic (E - P_\phic \gamma_\lambda)
\]
so the sign of $\ddot{\phic}$ at a bounce is correlated with the sign of
$E - P_\phic \gamma_\lambda$.

\section{Exiting Hagedorn with a two-derivative action\label{sect:inc}}

In this appendix we study the space of initial conditions which allows
the universe to start in an initial Hagedorn phase and subsequently
exit.  The discussion is based on the two-derivative effective action
$S_0$ given in (\ref{s0}).

\subsection{Requirements on initial conditions in the Hagedorn phase}
The starting point for the universe in string gas models is a high
temperature equilibrium phase near the self-dual radius. In this
initial Hagedorn phase, to a good approximation, the pressure vanishes
and the energy is constant. The assumption of equilibrium is quite
reasonable, as the interaction rates of strings are enhanced due to
their large oscillator numbers (lots of string in a small space). As
stated, the universe is taken to be near the self-dual radius,
$\lambda \approx 0$, and nearly static, $\dot{\lambda} \approx 0$.

Now let's consider fluctuations about this equilibrium configuration.
Suppose a thermal fluctuation allows $d$ dimensions to grow,
and for simplicity take this part of the universe to be isotropic.
The equations of motion become ($P_\lambda=P_\phic=0$)
\[\begin{aligned}
E&=(2\pi)^2e^{-\phic}(\dot{\phic}^2-d\dot{\lambda}^2)\\
\ddot{\phic}&=\frac{1}{2}(\dot{\phic}^2+d\dot{\lambda}^2)\\
\ddot{\lambda}&=\dot{\phic}\dot{\lambda}
\end{aligned}\]
With $E$ constant, these can be solved exactly, with
\[\begin{aligned}
\phic(t)&=\log\left[\frac{e^{\phic_0}}{(E/16\pi^2)e^{\phic_0}t^2-\dot{\phic}_0t+1}\right]\\
\lambda(t)&=A+\frac{1}{\sqrt{d}}\log\left[\frac{(E/8\pi^2)e^{\phic_0}t-(\dot{\phic}_0+\sqrt{d}\dot{\lambda}_0)}{(E/8\pi^2)e^{\phic_0}t-(\dot{\phic}_0-\sqrt{d}\dot{\lambda}_0)}\right]
\end{aligned}\]
The subscripts $0$ denote values at $t=0$ and are assumed to satisfy the Hamiltonian
constraint. In these solutions, the constant $A$ is the asymptotic value of the
scale factor, related to initial conditions by
\begin{equation}\label{ena}
A=\lambda(t\rightarrow\infty)=\lambda_0+\frac{1}{\sqrt{d}}\log\left[\frac{-(\dot{\phic}_0-\sqrt{d}\dot{\lambda}_0)}{-(\dot{\phic}_0+\sqrt{d}\dot{\lambda}_0)}\right]\,.
\end{equation}
(We are considering $\dot{\phic}_0<0$ and $\dot{\lambda}_0>0$, while for positive
matter energy we must have $\sqrt{d}\dot{\lambda}_0<|\dot{\phic}_0|$, so
the arguments of the logarithms are positive.)

At some radius larger than $\lambda_0$ the universe is expected to fall out of equilibrium and 
enter a new ``large radius'' phase. An exit condition, since the energy in the
Hagedorn phase is constant, can be expressed in the form
\begin{equation}\label{dio}\rho(\lambda_{\rm exit})=\rho_{H}\end{equation}
where 
\[\rho(\lambda)=\frac{Ee^{-d\lambda}}{(2\pi)^d}\]
is the energy density in the universe and $\rho_{H}$ is a characteristic
energy density of order $\alpha^{\prime-(\frac{d+1}{2})}$.  We will be more precise about
the value of $\rho_H$ below.
With the Hagedorn phase solution above, in order to exit to a radiation era we need 
\[\lambda(t\rightarrow\infty)>\lambda_{\rm exit}.\]
Using (\ref{ena}) and (\ref{dio}) this can be written
conveniently in terms of the variable
$x\equiv\frac{\sqrt{d}\dot{\lambda}_0}{|\dot{\phic}_0|}$ as
$x>(r-1)/(r+1)$
where $r\equiv\big(\rho(\lambda_0)/\rho_H\big)^{1/\sqrt{d}}$ is the ratio
of the initial energy density to the critical Hagedorn density. To be in the
Hagedorn era at $t=0$ we need $r>1$. Also we need $x<1$ for positive matter energy.

\subsection{Testing the phase space of initial conditions}
The equations of motion have four initial conditions, which can be chosen to
be the initial volume $V_0=(2\pi)^de^{d\lambda_0}$, $\phic_0$, $\dot{\phic}_0$ and
$x$ (instead of $\dot{\lambda}$, once $\dot{\phic}_0$ is fixed). The variable
$x$ reflects the initial ``boost'' of the scale factors. Fixing the first
three initial conditions, we then ask whether a given $x$ can drive the system out of the
Hagedorn phase.

The conditions on initial conditions such that the system starts in the Hagedorn era
($\rho(\lambda_0)/\rho_H>1$) and exits to the large radius era ($x > (r-1)/(r+1)$)
can be written as
\[\begin{aligned}
f_1&\equiv K_0(1-x^2)-1>0\\
f_2&\equiv
x-\frac{K_0^{1/\sqrt{d}}(1-x^2)^{1/\sqrt{d}}-1}{K_0^{1/\sqrt{d}}(1-x^2)^{1/\sqrt{d}}+1}>0\end{aligned}\]
where
\[K_0(\phic_0,\dot{\phic}_0,V_0)\equiv \frac{(2\pi)^2e^{-\phic_0}\dot{\phic}_0^2}{\rho_HV_0}.\]
In the Hagedorn phase the entropy in matter is to a good approximation
proportional to the energy. We use this in all that follows. With
$S=E/T_H$, the distribution of $x$ once the other three initial
conditions are fixed is a Gaussian,
\be
\label{deqn}
d(x)\sim e^{-(4\pi^2e^{-\phic_0}\dot{\phic}_0^2/T_H)x^2}\,.
\ee

Now let's be more precise about the condition for Hagedorn exit.  In the
Hagedorn phase the string gas is taken to be in equilibrium at temperature $T_H$.
This is similar to having a black hole in thermal equilibrium with the surrounding radiation,
something which can only happen in finite volume. The constraints on the
volume in which such equilibrium can be maintained have been
studied, and for a string gas in $d$ spatial dimensions read \cite{bow1,kounnas,hawking}
\begin{equation}
V_d<\frac{E}{c_d T_c^{d+1}}\,.
\end{equation}
Here $T_c$ is the temperature and
\begin{equation}
c_d=128\frac{2d!\zeta(d+1)}{(4\pi)^{d/2}\Gamma(d/2)}(2-2^{-d})
\end{equation}
is the Stefan-Boltzmann constant for 128 fermionic and 128 bosonic massless degrees of
freedom.  This suggests that
\begin{equation}\label{rcd}
\rho_H\simeq c_dT_H^{d+1}\equiv\rho_c(d)\,.
\end{equation}
Below this energy density the string gas will decay to radiation.

\begin{figure}[htp]
\centering
\includegraphics[scale=0.7,viewport=20 400 520 750,clip]{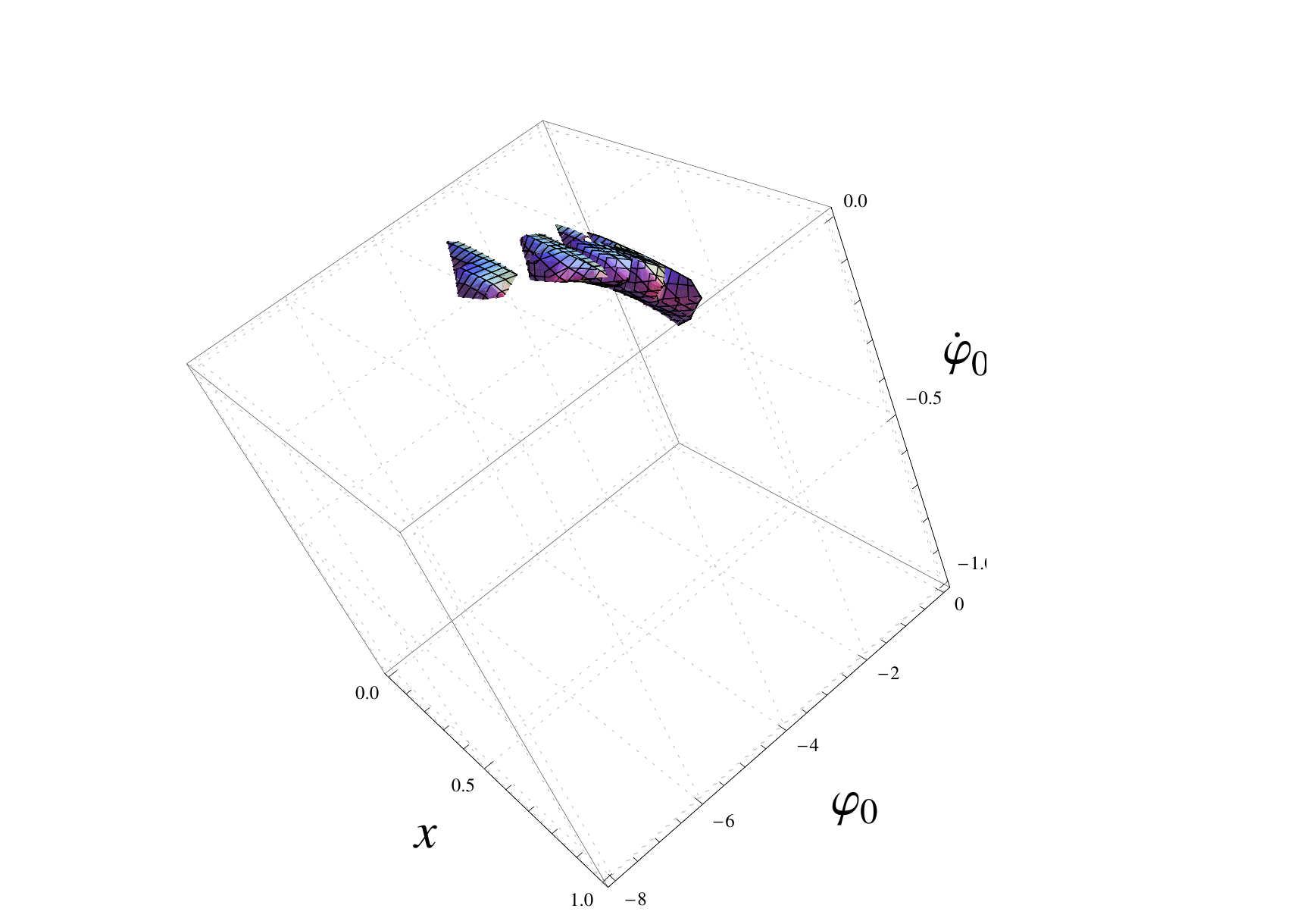}
\caption{{The shaded region is the space of initial
    conditions for which exit from the Hagedorn phase is possible. It is the
    region where $f_1$ and $f_2$ are positive and $x$ is within 3 standard deviations
    of its mean. We set $\rho_H=\rho_c(d)$ and $d=3$.}}\label{inc3d1}
\includegraphics[scale=0.85,viewport=20 500 520 750,clip]{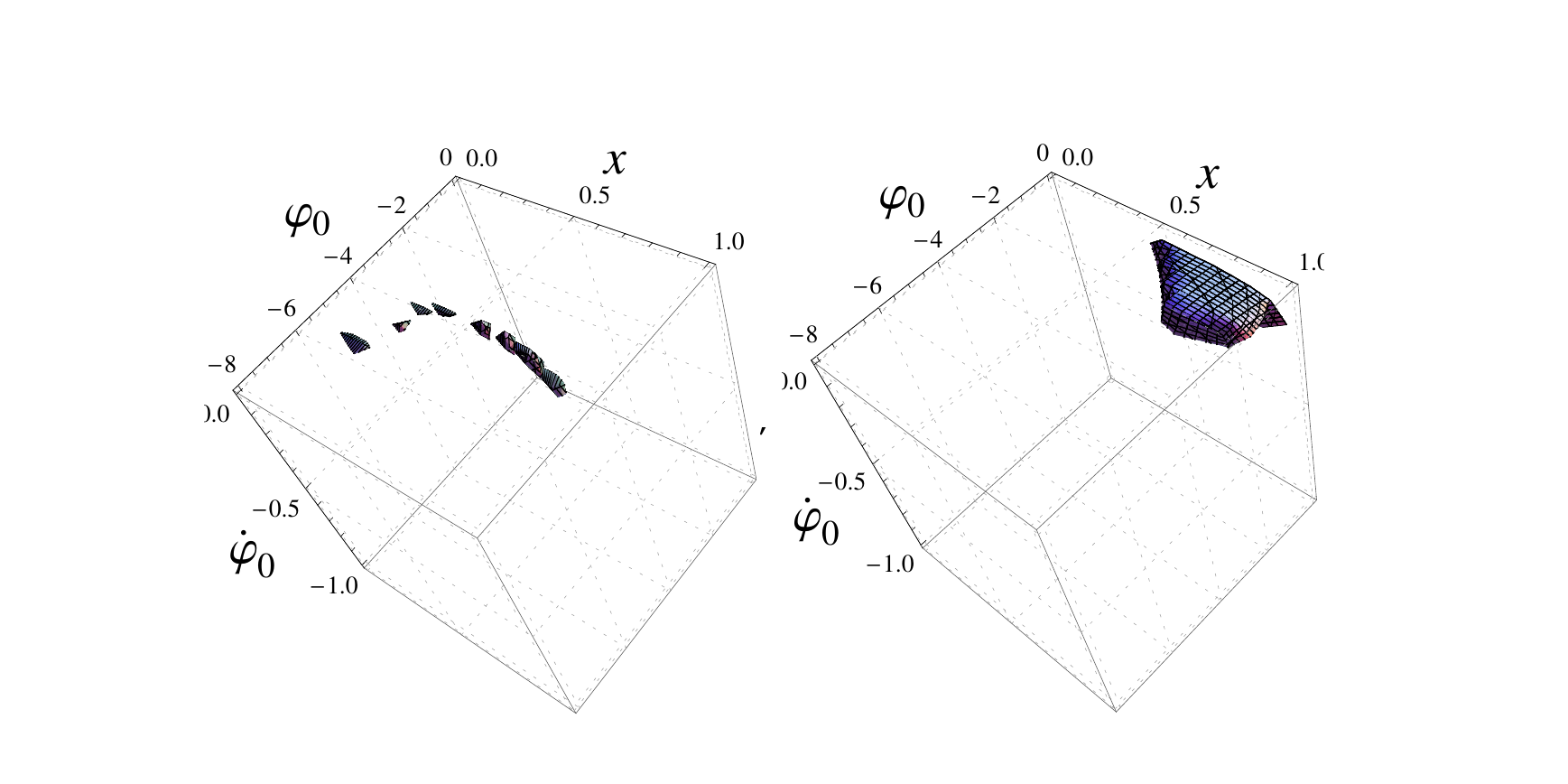}
\caption{{Same as figure \ref{inc3d1} but with
    $\rho_H=10\rho_c(d)$ (left) and $\rho_H=0.01\rho_c(d)$ (right).}} 
\label{inc3d23}
\end{figure}

To study the probability of exiting Hagedorn we fix the initial volume to be $V_0=(2\pi)^d$, that is,
we study fluctuations when the universe is at the self dual radius, and survey
the remaining 3-dimensional space of initial conditions. Figure \ref{inc3d1} shows the overlap of
the region where $f_1,f_2>0$ with the region where $x$ is taken to be
within 3 standard deviations of its mean in the distribution (\ref{deqn}).  The space of ``good''
initial conditions is obviously restricted. Figure \ref{inc3d23} is similar but with
$\rho_H=10\rho_c(d)$ and $\rho_H=0.01\rho_c(d)$.  The fact that the plots are similar shows that
our results are not sensitive to our estimate for $\rho_H$.  All the plots are for $d=3$, but they
change only slightly for different $d$.

One may still question whether there is indeed a problem with initial
conditions. A fluctuation to a large radius era, where winding modes
will want to annihilate and radiation will take over, could be rare
but if the system is kept in the Hagedorn phase such a fluctuation
will eventually occur. But this does not happen if one takes the
effective dilaton gravity action $S_0$ seriously: it implies that the
universe begins from an initial singularity and has only a finite
amount of time before the dilaton rolls to weak coupling.


\providecommand{\href}[2]{#2}\begingroup\raggedright\endgroup

\end{document}